\documentstyle[prl,aps,multicol,epsf]{revtex}
\begin{document} 
\draft 
\title
{Low energy excitations of the $t$$-$$J$ model in one
and two dimensions}

\author{R.Eder$^1$, Y.C. Chen$^2$, H.Q. Lin$^3$, Y. Ohta$^4$, C.T.
Shih$^5$, and T.K. Lee$^6$}
\address{
$^{1}$Dept. of Applied and Solid State Physics,
Univ. of Groningen, NL\\
$^{2}$Dept. of Physics, Tung-Hai Univ.,Taichung, Taiwan\\
$^{3}$Dept. of Physics, Chinese Univ. of Hong Kong, HK\\
$^{4}$Dept. of Physics, Chiba Univ., Chiba, Japan\\
$^{5}$Dept. of Physics, National Tsing Hua Univ., Hsinchu, Taiwan\\
$^{6}$Dept. of Physics, Virginia Tech, Blacksburg, VA;\\
Inst. of Physics, Academia Sinica, Nankang, Taipei 11529, Taiwan}
\maketitle

\begin{abstract} 
We present an exact diagonalization study of the low energy singlet
and triplet states for both $1D$ and $2D$ $t$$-$$J$ model.
A scan of the parameter ratio $J/t$ shows that for most
low energy states in both $1D$ and $2D$ the
excitation energy takes the form $E(t,J) = a\cdot t + b\cdot J$.
In  $1D$ this is the natural consequence of the factorization of the
low energy wave functions, i.e. spin-charge separation.
Examination of the low energy eigenstates in $2D$ shows that
most of these are collective modes,
which for larger $J$ correspond to a periodic modulation of the
hole density. The modulation is well reproduced by
treating holes as hard-core Bosons with an attractive interaction.
\end{abstract}
\begin{figure}
\epsfxsize=10.1cm
\vspace{-1.75cm}
\hspace{-0.5cm}\epsffile{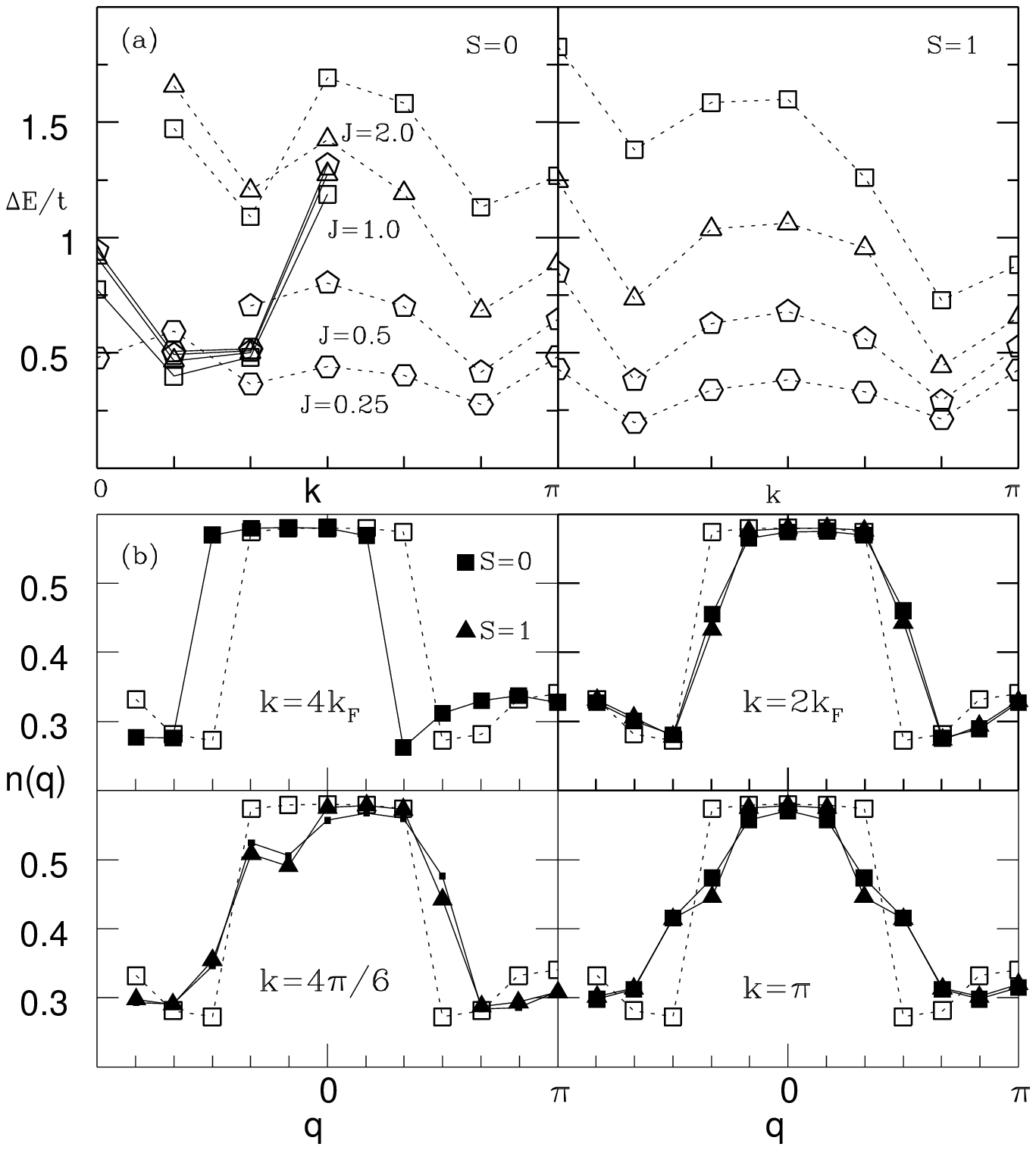}
\vspace{-1.0cm}
\narrowtext
\caption[]{(a) Lowest excitation energies for two holes in a $12$-site ring
for all possible momenta and different $J$. The figure shows the lowest or
lowest two states for each momentum, the lines are guide to the eye.\\
(b) EMD for the ground state 
of $2$ holes in the $12$-site ring (dashed line)
compared to $n(q)$ for some excited states with given
momentum $k$ (full lines). The ratio $J/t$$=$$0.5$.}
\label{fig1} 
\end{figure}

\begin{figure}
\epsfxsize=8.1cm
\vspace{-2.5cm}
\hspace{0ex}\epsffile{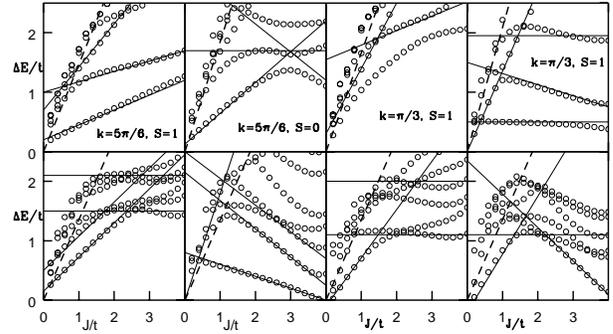}
\vspace{-1.3cm}
\narrowtext
\caption[]{Lowest excitation energies $\Delta E$ for the $12$ site ring
with $2$ holes (top row) and $4$ holes (bottom row). Spin and
momentum quantum numbers refer to both panels of one column.}
\label{fig2} 
\end{figure}

\begin{figure}
\epsfxsize=10cm
\vspace{-0.5cm}
\hspace{-1cm}\epsffile{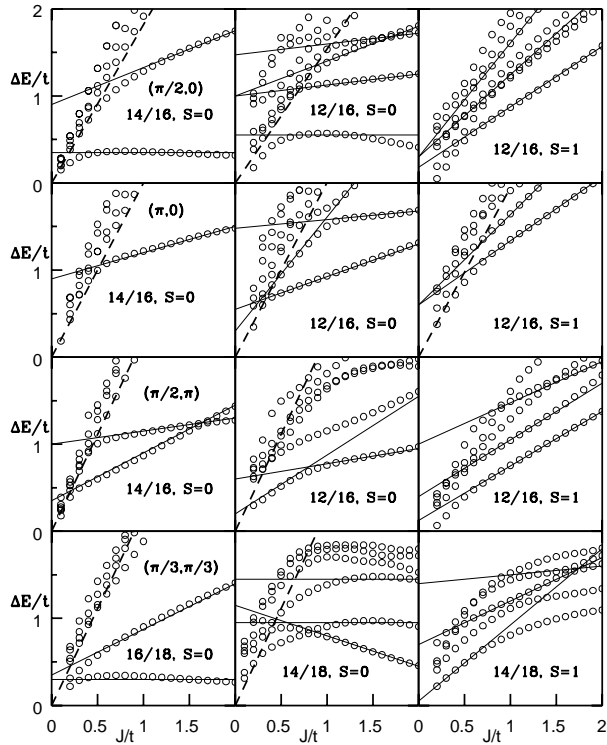}
\vspace{0.5cm}
\narrowtext
\caption[]{Lowest excitation energies $\Delta E$ for different 
$2D$ cluster sizes $N$ and electron numbers $N_e$ in $2D$.
Each row corresponds to one momentum, the graphs are labelled by
electron concentration $N_e/N$ and spin $S$.}
\label{fig3} 
\end{figure}

\begin{figure}
\epsfxsize=9.1cm
\vspace{-0.75cm}
\hspace{0ex}\epsffile{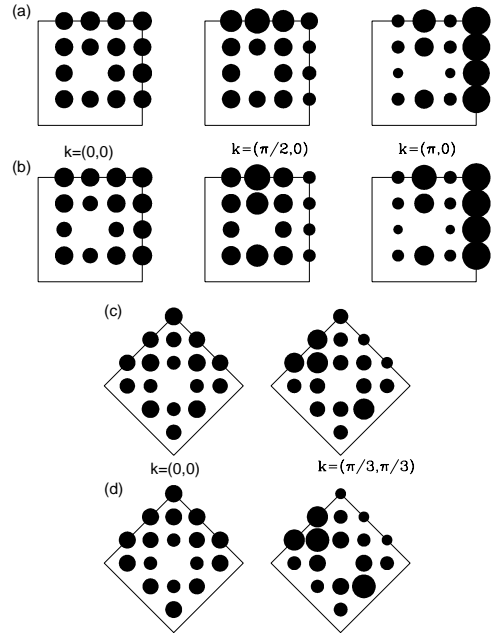}
\vspace{0cm}
\narrowtext
\caption[]{Hole-density correlation function $g(\bbox{R})$
for the lowest $S$$=$$0$ eigenstates with $4$ holes/hard core bosons
and different total momentum $\bbox{k}$:
(a) $t$$-$$J$ model, $16$ sites,  (b) hard core bosons,
$16$ sites,  (c) $t$$-$$J$ model, $18$ sites,
(c) hard core bosons, $18$ sites.
The radius of the circle at each $\bbox{R}$ is proportional to 
the value of $g(\bbox{R})$,
the ratio $J/t$$=$$1$ for the $t$$-$$J$ model and
$V/t$$=$$1$ for the bosons.}
\label{fig4} 
\end{figure}

\begin{figure}
\epsfxsize=10.1cm
\vspace{-0.5cm}
\hspace{-1cm}\epsffile{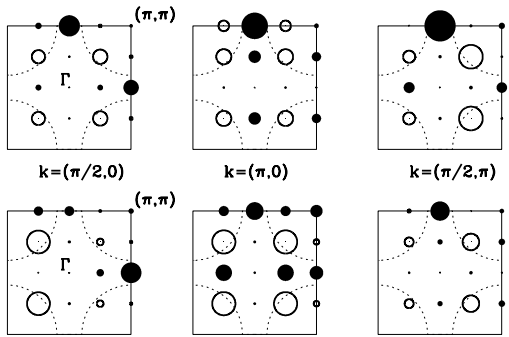}
\vspace{-4.5cm}
\narrowtext
\caption[]{
Difference $\Delta n(\bbox{q})$ for various low energy states
in the $4\times 4$ cluster with $4$ holes. For each
$\bbox{q}$ an open (full) circle
indicate an positive (negative) $\Delta n(\bbox{q})$, the radius
of the circle is proportional to the magnitude.
The ratio $J/t$$=$$0.5$, the $\Delta n(\bbox{k})$ refer to the
lowest singlet (top row) and triplet (bottom row) state
with momentum $\bbox{k}$.}
\label{fig5} 
\end{figure}

 
\begin{multicols}{2}
Deviations from `generic' Fermi liquid behavior observed in the normal
state of high-temperature superconductors have inspired a
search for exotic ground states of strongly correlated electron
systems. Thereby a $2$ dimensional
version of the Tomonaga-Luttinger liquid (TLL), which
is realized in various $1D$ systems\cite{OgataShiba,BaresBlatter},
has received considerable attention. In a TLL the entire low energy
excitation spectrum consists of collective excitations,
resulting in a `decay' of the physical electron into
pairs of such collective modes and consequently rather unusual physics.\\
Exact diagonalization 
allows to systematically scan the entire low energy
spectrum of small clusters and even to analyze the wave 
functions of low energy states and thus provides a unique tool to obtain
approximation-free, unbiased results for $2D$ strong correlation systems.
While the effects of the cluster size
pose a serious limitation,
it will be shown below that already quite small systems
suffice to clearly demonstrate e.g. 
the characteristic features of the TLL in $1D$.
This suggests to seek analogies between rings and planar clusters,
and, as will be shown below, $2D$ systems indeed show 
some similarity with $1D$. In particular,
in both cases we find a very regular scaling behavior of the
lowest excitation energies as well as
evidence for a large number of low energy collective excitations.
The collective charge excitations can be modelled by
treating holes as hard-core Bosons.\\
We begin with a discussion of $1D$. Figure \ref{fig1}a 
shows the dispersion of the excitation
energy $\Delta E$ (i.e the energy relative to the ground state energy)
for the lowermost singlet and triplet states of a $12$-site ring
with two holes. Results for both longer and 
shorter rings (down to $6$ sites) are completely consistent with these.
In the $S$$=$$0$ sector one can identify
two `branches' of states with very different dependence
on $J$: a first branch (connected by dashed lines) 
resembles the lower edge of a
particle-hole continuum, and has its lowest excitation energy at
$2k_F$$=$$5\pi/6$. The excitation energy roughly scales with $J$, although
\noindent the energies do not really scale to zero as $J$$\rightarrow$$0$
(which however is likely to be a finite-size effect). 
A second branch (connected by a full line)
comes down at $4k_F$$=$$2\pi/6$,
and  is practically invariant under a change of $J$, i.e. its
energy scale is $t$. 
The lowest triplet states closely resemble the dispersion
and scaling properties of the singlet $J$-branch, having their
minimum excitation energy at $2k_F$. 
The singlet $t$-branch and triplet $J$-branch have been discussed
by Bares and Blatter based on the Bethe-ansatz solution at the
supersymmetric point, $J$$=$$2t$\cite{BaresBlatter}; the (near) identity
of low energy singlet and triplet spectrum is known
for the undoped $1D$ Heisenberg antiferromagnet, where a `singlet replica'
of the de Cloiseaux-Pearson mode has been discussed by 
Mueller {\em et al.}\cite{Mueller}. This suggests that
in the language of the Ogata-Shiba wave function\cite{OgataShiba} 
the $J$-branches
correspond to excitations of the spin part of the wave function,
the $t$-branch to an excitation of spinless Fermions 
or, equivalently, hard-core Bosons.\\
The unusual nature of the low energy excitations can be nicely seen in
the electron momentum distribution (EMD), defined as
$n(q) = \langle c_{q,\uparrow}^\dagger
c_{q,\uparrow} \rangle$. This is shown in Figure 
\ref{fig1}b for some of the low energy states with $S_z$$=$$0$.
The EMDs of the excited states differ appreciably 
from that of the ground state only near $\pm k_F$, which
is in principle what one would expect for particle-hole
excitations in a Fermi liquid.
The only excited states, however, where the change of $n(q)$ 
would be consistent with the Fermi-liquid picture are the ones with 
momentum $k$$=$$2k_F$. Here the decrease of $n(q)$ 
at $q$$=$$-2\pi/6$ and the nearly equal increase at 
$q$$=$$3\pi/6$ seem to indicate the shift of an electron
between these two momenta, as expected for a particle-hole excitation with 
a momentum transfer of $k_F$$=$$5\pi/6$. For the excited
states with other momenta no such assignment of an electron
transfer with the proper momentum is possible
(for example the $4k_F$ singlet
state `looks like' a particle-hole excitation
$2\pi/6$$\rightarrow$$-3\pi/6$, which however would give
a wrong momentum transfer of $-5\pi/6$ rather than $2\pi/6$).
These results may be understood by recalling that
the low energy excited states may be thought of\cite{BaresBlatter} as
particle-hole excitations of either a spinon or a holon,
whereas an electron near $k_F$ corresponds to a combination of
low energy spinon-holon pairs. Changing
the `occupation numbers' of spinons and holons
therefore will in turn change the $n(q)$
of the electrons near $k_F$. How $n(q)$ changes in detail,
however, depends on the (unknown) `expansion' of the
physical electrons in terms of spinon-holon pairs, and thus need not
at all be consistent with the particle-hole picture.\\
While the lowermost states correspond to `pure' excitations
of either spin or charge degrees of freedom, there is no reason
why there should be no `mixed' excitations,
where spin and charge part of the wave function are excited
simultaneously.
This is demonstrated in
Figure \ref{fig2}, which shows the excitation energies,
for a larger part 
of the low energy spectrum for both $2$ and $4$ holes
in the $12$-site ring.
Some of the excitation energies (indicated by full lines) have
the form
$\Delta E(t,J)$$ =$$ a\cdot t$$ +$$ b\cdot J$. 
This could be understood if one were to assume that for these states both,
charge and spin degrees of freedom are
excited, the former having $t$ as their energy scale, the latter having
$J$. 
When two or more of the linear $\Delta E$ vs. $J$ curves intersect,
they sometimes produce pronounced `hybridization gaps'; we believe
this is the case e.g.
in the singlet sector for two holes and momentum $5\pi/6$.
\noindent It should be noted that states in one panel 
are identical in all their possible quantum numbers; 
nonvanishing matrix elements of the Hamiltonian are therefore
possible, and depending
on the strength of these matrix elements
levels may `repel' each other appreciably. 
In addition, there appear to be deviations
from the `ideal' behavior for large $J$; we believe that this
is due to the increasing tendency towards hole clustering, which
in the thermodynamic limit is known to lead to phase separation
for $J$ larger than $J_c$$\approx$$3.5$\cite{Ogataetal}. In the small
clusters there is no true phase transition, but rather a very
slow and gradual crossover to the clustered state, which
may lead to a `bending' of the $\Delta E$ vs. $J$ curves.
The level crossing at $J/t$$\approx$$4$, where the $4$ hole singlet state
with $k$$=$$5\pi/6$ comes down below the ground state,
most probably is also related to hole clustering.
Summarizing the results in $1D$, the
energies of most of the low lying eigenstates to a good approximation
can be written in the form 
$\Delta E(t,J)$$ =$$ a\cdot t$$ +$$ b\cdot J$. 
This can be understood as a consequence of the factorization
of the eigenfunctions into a spin part and
a spinless Fermion part, as established by
Ogata and Shiba in the limit $J/t\rightarrow 0$.\\
Motivated by these systematics we have performed a scan of the $J/t$
dependence of the lowermost excitation energies in $2D$.
In $2D$ point group symmetry presents
a slight complication, in that all allowed momenta
for both the $16$ and $18$ sites clusters are along high symmetry lines;
the respective eigenstates then can be classified according to their
parity under reflection by the high-symmetry line.
In the following we have
only selected those states, whose parity equals that of the
ground state, so that they could in principle be seen
in the dynamical charge correlation function (DCF).
Then, Figure \ref{fig3} shows some representative
data for the $J$ dependence of the lowest singlet
and triplet excitation energies.
As was the case for $1D$ some of the excitation energies
(indicated by full lines) have 
the form $\Delta E(t,J) = a\cdot t + b\cdot J$. As 
in $1D$ there are dense continua (indicated by dashed lines)
with energy scale $J$ coming down at small $J$
so that the linear branches `get submerged'.
For some of the states the $\Delta E$ vs. $J$ curve has a
negative curvature; we believe that this is a manifestation
of the stronger tendency towards hole clustering in $2D$. 
Extending the range of $J$ up to
extreme values such as $J/t$$=$$10$, where the energy
is essentially determined by the number of broken bonds, the energies
of these states approach that of the ground state. We therefore
believe that the negative curvature is a consequence of increasing
hole clustering, which in the thermodynamic limit
leads to phase separation for $J$ greater than 
$J_c$$\approx>$$1.5$\cite{Puttikaetal}. 
Since this
critical value is smaller in $2D$ than in $1D$
it seems plausible that in $2D$ the excitation energies
are influenced stronger. Another 
\noindent noteworthy point is that
previous studies\cite{EderOhtaMaekawa} indicate that
most of the states which contribute significantly in the DCF
seem to be `pure $t$-branches'.\\
Despite some complications
the above data show an undeniable analogy between the low 
energy spectra of $1D$ and $2D$. 
To find out more about the nature of the
$2D$ eigenstates, we proceed to an analysis of their
wave functions. We start with a discussion of the static
hole density correlation function, defined as
\[
g(\bbox{R}) = 
\sum_i \langle ( 1- n_i)(1-n_{i+\bbox{R}}) \rangle,
\]
where $n_i$ is the operator of electron density at site $i$.
The expectation value is taken with the respective
eigenstate, some representative results are given 
in Figure \ref{fig4}. In the ground state (with momentum
$(0,0)$) for $J/t$$=$$1$, $g(\bbox{R})$ is uniform
in $16$ sites and even shows a slight decrease at small distances
in $18$ sites. Contrary to the fairly uniform
density correlation in the ground states, the low energy
\noindent singlet states show pronounced and
systematic modulations of $g(\bbox{R})$.
In particular, for momentum along $(1,0)$ these states
seem to correspond to `dynamical hole columns'.
This is reminiscent of the results of Prelovsek and 
Zotos\cite{PrelovsekZotos}, who
studied static four-point correlation functions and 
for large enough $J/t$ found
indications of column-like hole correlations in the ground state.
Our scan of the low energy spectrum indicates that
the respective states develop quite continuously down to small
values of $J/t$ (for example the state at $\bbox{k}$$=$$(\pi,0)$
has a completely linear $\Delta E$ down to $J$$=$$0.5$);
in this physical parameter regime, however, the column-like structures 
become nearly unobservable. In previous studies\cite{Zotos,Chen1},
the correlation of holes in the ground state
was found to be very similar to that of hard-core Bosons.
By adding an additional nearest-neighbor attractive
interaction $V$ between hard-core Bosons we could
indeed reproduce the $g(\bbox{R})$ of the
excited states of the $t$$-$$J$ model
surprisingly well, as shown in Figure \ref{fig4}. 
We believe that this is an indication that the columns of holes
represent a sound-like excitation which gains additional
stabilization by a residual short range attractive interaction between
holes.
We also note that
such column-like structures are not obtained when we excite a
particle-hole pair from the projected ideal Fermi gas.\\
The collective nature of the low lying states again
becomes very obvious in the EMD.
More precisely, we consider the difference $\Delta n(\bbox{q})$
of the EMD for an excited state minus 
the EMD of the ground state; this quantity reflects
the redistribution of electrons in momentum space when
going over from the ground state to the excited state.
Figure \ref{fig5} shows $\Delta n(\bbox{q})$ for some of the
low energy states of the $16$ site cluster.
As was the case in $1D$, the changes of $n(\bbox{q})$ occur predominantly
in the region of $\bbox{k}$ space where the EMD of the ground state
drops 
\noindent sharply; this region is indicated by the dashed
line in Figure \ref{fig5}.
Moreover, it is usually
impossible to assign transitions of a single electron 
which could explain the momentum of the state. 
Let us  for example consider the $\Delta n(\bbox{q})$ for
the triplet state with $\bbox{k}=(\pi/2,0)$. Here it might seem
that an electron has been moved from
$(\pi,0)\rightarrow (-\pi/2,\pm \pi/2)$. However, this is
inconsistent
with the momentum of the state, $(\pi/2,0)$,
so that this state obviously cannot correspond to a single
particle-hole excitation. We have performed a similar analysis
for other momenta
and point group symmetries, as well as for $6$ holes in $16$ sites
for both $t$$-$$J$ and Hubbard model\cite{Chen},
and found consistency with the particle-hole picture only in few cases.
The situation is very
different at lower electron concentration, where
Fermi-liquid like particle-hole transitions can be clearly
identified in the EMD for all low energy states\cite{EderOhta}.
Then, summarizing the numerical results for $2D$,  
there is first of all an obvious analogy with $1D$.
Many of the low lying excitation energies
have a parameter dependence of the form
$\Delta E(t,J)$$ =$$ a\cdot t$$ +$$ b\cdot J$,
which in $1D$ reflects the
factorization of the low lying eigenfunctions.
As was the case in $1D$
most of the low energy eigenstates in $2D$
seem to be collective modes, corresponding
to excitations of several different particle-hole pairs.
For large enough $J$ some of these
develop into `dynamical hole lines'. 
This may partly explain the analogy with $1D$,
in that a line of holes would `cut' the spin background
into two parts in much the same way as a single hole would do 
for a $1D$ system. The lowest of these column-like states can be
followed from the large $J$ limit down to values
as small as $J$$\approx$$0.5-0.3$, and have relatively small excitation
energies. It should be noted that our clusters 
allow only for density modulations that are
`harmonics' of the cluster boundary
and that finite-size effects
most probably tend to increase excitation energies.
In an infinite system these collective modes therefore may well appear
at low energies and in the physical regime of parameters.\\
As for the importance of these collective modes for the low energy physics
one could envisage different scenarios. It might be that they
simply represent an  additional complication
`on top of' a simple Fermi liquid-like state, i.e. a kind of
sound waves. 
A second possibility would be that, similar as in
$1D$, the collective modes comprise the entire low energy
spectrum, leading to the well-known `decay' of the physical
electron, and thus some $2D$ analogue of the
Tomonaga-Luttinger liquid. 
Finally, if one were to assume that extremely low energetic
`hole column' excitations persist in some way in the physical 
parameter range and that these columns' zero-point fluctuations
dominate the low energy physics, one would arrive at the
`meandering domain wall fluid'\cite{Zaanen}.
However, for
any of these scenarios,  the fact that
the low-energy collective charge
excitations also have density correlations similar to hard-core
Bosons  suggest a non-conventional metallic state
in this system.\\
Financial support of R. E. by the European community
is most gratefully acknowledged.
Part of the research was conducted using the resources of the Cornell Theory
Center and National Center for High Performance Computing in Taiwan. 
 
\end{multicols}
\end{document}